\documentclass[aps,twocolumn,tightenlines,amsmath,amssymb,superscriptaddress]{revtex4-2}

\usepackage{graphicx}
\usepackage{amsmath}
\usepackage{amssymb}
\usepackage{epstopdf}
\usepackage{color}
\usepackage{lipsum}

\usepackage[colorlinks=true]{hyperref}
\hypersetup{citecolor=blue}

\usepackage{ulem}
\usepackage[dvipsnames]{xcolor}
\usepackage{todonotes}

\DeclareGraphicsExtensions{.eps}

\begin{document}
\title{Photonic source of heralded GHZ states}

\author{H.~Cao}
\affiliation{University of Vienna, Faculty of Physics, Vienna Center for Quantum Science and Technology (VCQ), 1090 Vienna, Austria}
\affiliation{Christian Doppler Laboratory for Photonic Quantum Computer, Faculty of Physics, University of Vienna, 1090 Vienna, Austria}
\author{L.~M.~Hansen}
\affiliation{University of Vienna, Faculty of Physics, Vienna Center for Quantum Science and Technology (VCQ), 1090 Vienna, Austria}
\affiliation{Christian Doppler Laboratory for Photonic Quantum Computer, Faculty of Physics, University of Vienna, 1090 Vienna, Austria}
\author{F.~Giorgino}
\affiliation{University of Vienna, Faculty of Physics, Vienna Center for Quantum Science and Technology (VCQ), 1090 Vienna, Austria}
\affiliation{Christian Doppler Laboratory for Photonic Quantum Computer, Faculty of Physics, University of Vienna, 1090 Vienna, Austria}
\author{L.~Carosini}
\affiliation{University of Vienna, Faculty of Physics, Vienna Center for Quantum Science and Technology (VCQ), 1090 Vienna, Austria}
\affiliation{Christian Doppler Laboratory for Photonic Quantum Computer, Faculty of Physics, University of Vienna, 1090 Vienna, Austria}
\author{P.~Zah\'alka}
\affiliation{Christian Doppler Laboratory for Photonic Quantum Computer, Faculty of Physics, University of Vienna, 1090 Vienna, Austria}
\author{F.~Zilk}
\affiliation{Christian Doppler Laboratory for Photonic Quantum Computer, Faculty of Physics, University of Vienna, 1090 Vienna, Austria}
\author{J.~C.~Loredo}
\email{juan.loredo@univie.ac.at}
\affiliation{University of Vienna, Faculty of Physics, Vienna Center for Quantum Science and Technology (VCQ), 1090 Vienna, Austria}
\affiliation{Christian Doppler Laboratory for Photonic Quantum Computer, Faculty of Physics, University of Vienna, 1090 Vienna, Austria}
\author{P.~Walther}
\email{philip.walther@univie.ac.at}
\affiliation{University of Vienna, Faculty of Physics, Vienna Center for Quantum Science and Technology (VCQ), 1090 Vienna, Austria}
\affiliation{Christian Doppler Laboratory for Photonic Quantum Computer, Faculty of Physics, University of Vienna, 1090 Vienna, Austria}

\begin{abstract}
Generating large multiphoton entangled states is of main interest due to enabling universal photonic quantum computing and all-optical quantum repeater nodes. These applications exploit measurement-based quantum computation using cluster states. Remarkably, it was shown that photonic cluster states of arbitrary size can be generated by using feasible heralded linear optics fusion gates that act on heralded three-photon Greenberger-Horne-Zeilinger (GHZ) states as the initial resource state. Thus, the capability of generating heralded GHZ states is of great importance for scaling up photonic quantum computing. Here, we experimentally demonstrate this required building block by reporting a polarisation-encoded heralded GHZ state of three photons, for which we build a high-rate six-photon source {($547{\pm}2$~Hz)} from a solid-state quantum emitter and a stable polarisation-based interferometer. The detection of three ancillary photons heralds the generation of three-photon GHZ states among the remaining particles with fidelities up to {$F{=}0.7278{\pm}0.0106$}. Our results initiate a path for scalable entangling operations using heralded linear-optics implementations.
\end{abstract}
\maketitle

\textit{Introduction.---}Quantum entanglement enables the exploration of unique phenomena absent in the classical world, such as non-locality~\cite{GHZ:Pan2000,nonlocal:Winter07,nonlocal:Wehner14} and teleportation~\cite{teleport:Wootters93,teleport:97,teleport:Pan15}; and it ultimately provides an advantage to quantum systems over classical ones for various tasks~\cite{QAReview:Deutsch20,QAreview:22}, ranging from metrology and sensing~\cite{QMetrology:11,QMetrology:Atoms18,QSensingRev:17,QSensingRev:18,QSensing:Filho18}, to computation~\cite{QSup:Martinis19,zhong2020quantum,QA:Lavoie22,QASupCirc:23}. Photonic quantum systems are among the leading physical platforms for large-scale quantum computers~\cite{LOQC:Milburn07}. A very promising architecture is based on the concept of measurement-based quantum computing (MBQC) that exploits entangled cluster states as resource for universal quantum  computing. A key advantage of this scheme is the superior feasibility and error-correction thresholds~\cite{LOQC:Rudolph05,Percolation:Eisert07,pant2019percolation} with respect to gate-based models. For generating photonic cluster states two main encoding schemes exist: analog or continuous variable~\cite{PhotonicsCV:Loock05,QCCV:Nielsen06}. and digitial or discrete-variables approaches~\cite{PhotonicRev:Sciarrino18,PhotonicRev:Pryde19}. With advantages and disadvantages from both sides, continuous-variable implementations are highly sensitive to losses, degrading the quality of the quantum state. In this regard, discrete-variable encoding constitutes an appealing alternative, as even in the absence of deterministic entangling gates, there exist loss-tolerant schemes for the generation of large entangled quantum states using only probabilistic, but heralded, linear-optics quantum gate operations~\cite{SLOQC:Rudolph08,3ghzQC:Rudolph15}.

Additional approaches exist that aim to generate discrete-variable cluster states~\cite{Clusters:Briegel01} directly by exploiting quantum emitters~\cite{LinCS:Rudolph09,LinCS:Economou19,LinCS:Economou22,LinCS:Economou23,detCS:Schwartz16,detCS:Rempe22}. However, this requires complex control of atomic structures, solid-state materials, and electro-magnetic fields, to name a few technological challenges. In contrast, linear-optics alone also provides a path for the scalable generation of multiphoton cluster states universal for quantum computation~\cite{LOQC:Rudolph05,EffLOQC:Rudolph05,3ghzQC:Rudolph15}.
Thereon, a ballistic (without feed-forward requirements) and loss-tolerant (where losses do not induce logical errors) model for universal quantum computing, named fusion-based quantum computation~\cite{FusionQC:Sparrow23}, exploits small resource states made up of a handful of entangled particles~\cite{Percolation:Eisert07}, and combines them into larger entangled states via boosted (heralded, and at the expense of ancillary photons) entangling gates called fusion operations~\cite{LOQC:Rudolph05}.%\ch{Thererefore, a scalable and loss-tolerant MBQC can be envisaged via the fusion-based quantum computation.}

The smallest building block in these protocols~\cite{3ghzQC:Rudolph15} is the heralded three-photon Greenberger-Horne-Zeilinger (GHZ) state~\cite{ghz:89,3ghz:Z99}. Creating them requires the quantum interference of six separable single photons~\cite{SLOQC:Rudolph08}, or a minimum of ten photons from non-linear frequency conversion processes~\cite{hGHZ:Zeilinger07,heraldedGHZ:Guo09}. The efficient generation of the necessary input multiphoton states remained to date a major challenge. In this regard, semiconductor quantum dots have recently matured to a point where one handles the interference of single photons at scales of eight particles~\cite{Multiphoton:Carosini23} and even beyond with lost-photons boosted protocols~\cite{BS20:Pan19}, thus now reaching the necessary scales for these more advanced experiments.

Here, we experimentally demonstrate a heralded three-photon polarisation-encoded GHZ state based on the interference of six single photons. We employ a {28.7$\%$ fibre-efficient} quantum dot single-photon source, actively demultiplexed to produce a source of six indistinguishable photons {with $547{\pm}2$~Hz} detected rates. The high quality of the source and interferometric apparatus enable producing heralded three-photon GHZ states at a detection rate of {$0.914{\pm}0.006$~Hz}, and fidelities up to {$\mathcal{F}{=}0.7278{\pm}0.0106$}. Our results mark an important step for enabling the realisation of future fusion-based quantum computing protocols.

	\begin{figure}[htp!]
		\centering
		\includegraphics[width=.4\textwidth]{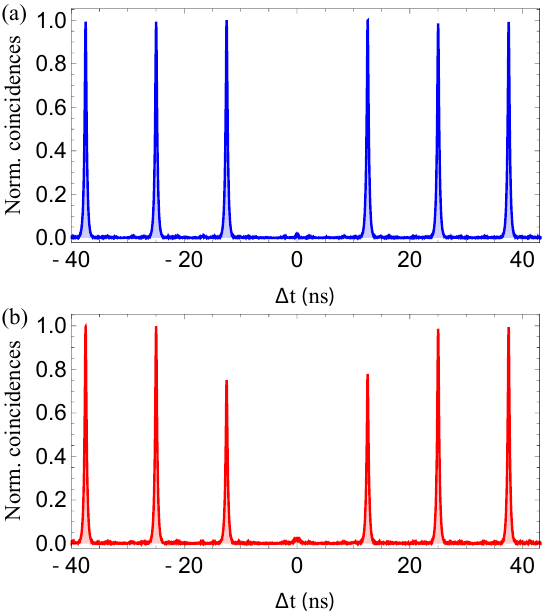}\vspace{0mm}
		\caption{\textbf{Single-photon quality.} (a) Normalised second-order auto-correlation function $g^{(2)}(\Delta t)$ from a Hanbury Brown and Twiss setup, and (b) at the output of a Hong-Ou-Mandel experiment at $\pi$-pulse excitation. We measure the single-photon purity {$1{-}g^{(2)}(0){=}0.981{\pm}0.003$}, and two-photon indistinguishability {$\mathcal{I}{=}0.941{\pm}0.002$}.}
	\label{fig:1}
	\end{figure}
	
	\begin{figure}[htp!]
		\centering
		\includegraphics[width=.48\textwidth]{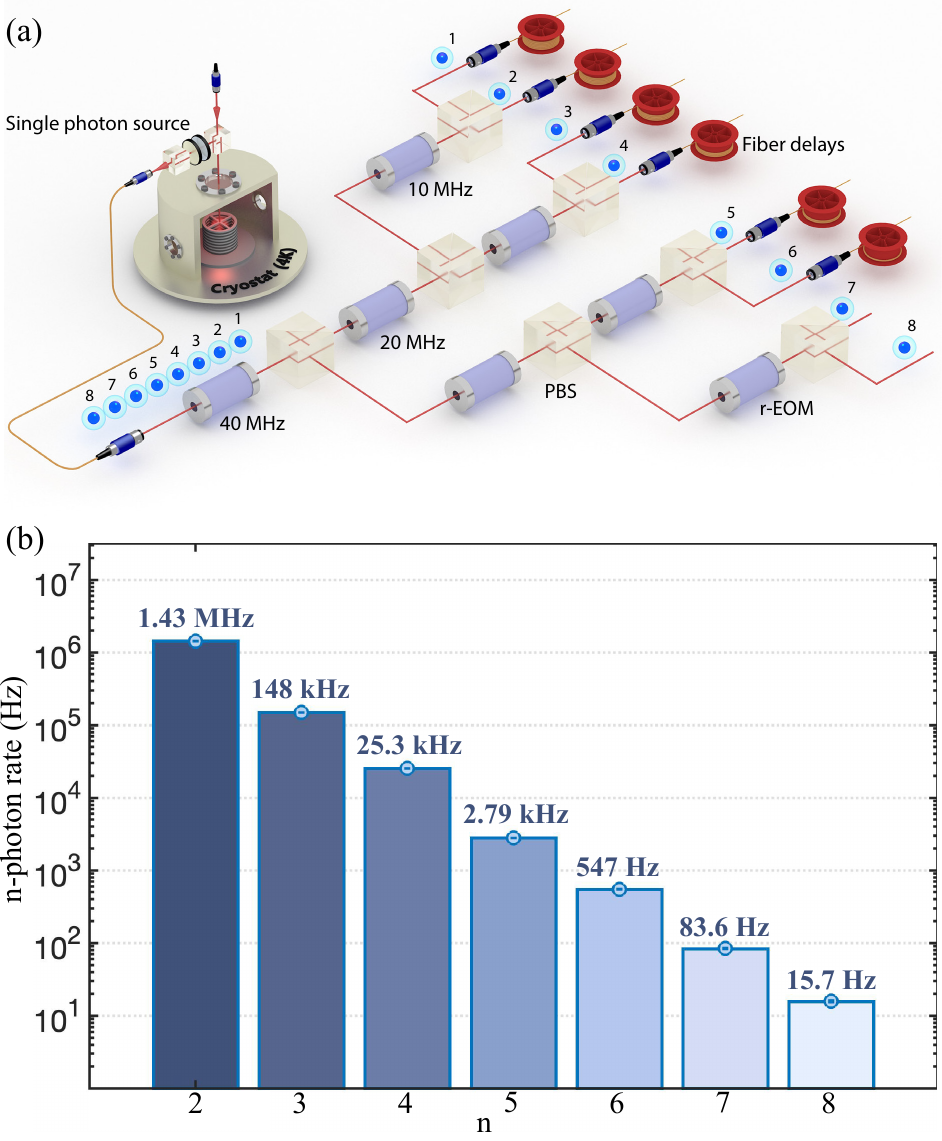}\vspace{0mm}
		\caption{\textbf{Multiphoton source.} (a) Resonant demultiplexer. Seven synchronised r-EOMs----one driven at $40$~MHz, two at $20$~MHz, and four at $10$~MHz--and polarising beam splitters, deterministically demultiplex eight consecutive time bins. Fibre-based delays and translation stages are used to correct the time bins' initial temporal mismatch. (b) Measured coincidence rates. Multiphoton rates at the output of the demultiplexer, up to a number of $n{=}8$ photons.}
	\label{fig:2}
	\end{figure}

\textit{High-rate multiphoton source.---}We first describe our source of multiphoton states. An InAs/GaAs quantum dot coupled to a micropillar cavity is kept in a cryostat at ${\sim}4$~K, and is resonantly driven using a standard crossed-polarised excitation scheme. An efficient collection setup allows us to measure {19.5~MHz} of single photons with simultaneous purity {$1{-}g^{2}(0){=}0.981{\pm}0.003$ and indistinguishability~\cite{HOM:Ollivier21} $\mathcal{I}{=}0.941{\pm}0.002$}, see Fig.~\ref{fig:1}, when pumped with $\pi$-pulses at 80~MHz repetition rate and using a detection system of {$85\%$ efficiency}, thus corresponding to a {$28.7\%$ fibre-efficient single-photon source}.

Subsequently, we utilise a time-to-space demultiplexer composed of resonant-enhanced electro-optic modulators (r-EOMs) and polarising beamsplitters (PBSs) arranged in a tree-structure~\cite{InterfacingSPP:19,demux:Keil22}, see Fig.~\ref{fig:2}(a). As a result, a number of input time bins separated from each other by 12.5~ns are deterministically routed towards, in this implementation, eight different outputs, which thereon follow suitable fibre-based temporal delays to result in a source of eight indistinguishable single photons travelling simultaneously. Figure~\ref{fig:2}(b) shows the measured multiphoton coincidence rates using eight superconducting nanowire single-photon detectors (SNSPDs) directly at the output of the demultiplexed source. In particular, the resulting six-photon source is detected at a rate of {$547{\pm}2$~Hz}, and the eight-photon source at {$15.7{\pm}0.4$~Hz.} We note that these are the highest rates of multiphoton sources reported to date.

\begin{figure*}[htp!]
		\centering
		\includegraphics[width=.999\textwidth]{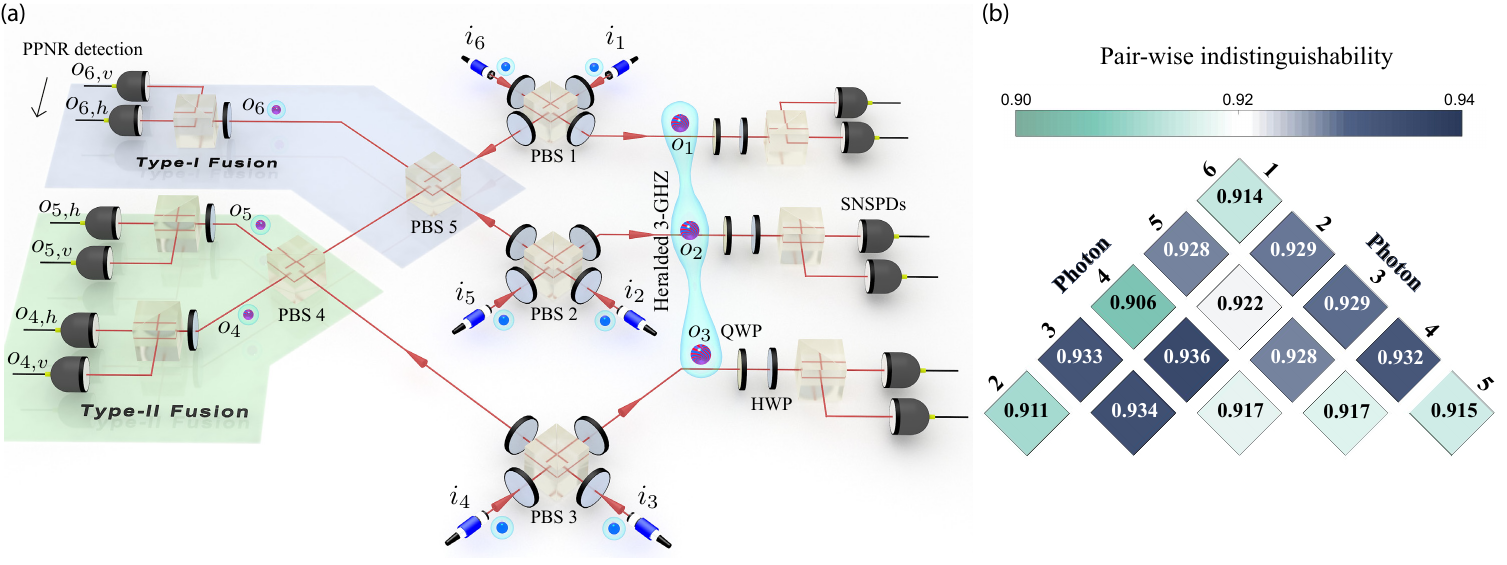}\vspace{0mm}
		\caption{\textbf{Polarisation six-photon interferometer.} (a) Depiction of experimental setup. Three pairs of single photons $\left(i_1,i_6\right), \left(i_2,i_5\right), \left(i_3,i_4\right)$ probabilistically generate three bell pairs, together with other unwanted terms, after interfering on PBSs 1,2,3. Subsequently, a non-heralded six-photon entangled state, and further unwanted states, is probabilistically generated after PBSs 4,5. The polarisation and number-resolved detection at the output of the type I and type II fusion operation among outputs $o_4, o_5$, and $o_6$ corrects for the unwanted terms and leaves the remaining photons at outputs $o_1, o_2$, and $o_3$ in a probabilistic but heralded three-photon GHZ state. Quater-wave (QWP) and half-wave plates (HWP) together with extra PBSs are used to perform three-qubit quantum state tomography. (b) Photons' indistinguishability. We use the same six-photon interferometer to measure all 15 pair-wise two-photon indistinguishabilities, resulting in an average value of {$0.923{\pm}0.009$} across all combinations.}
	\label{fig:3}
	\end{figure*}

\textit{Heralded entanglement.---}We use six single photons from this source as input to a polarisation-based interferometer, as depicted in Fig.~\ref{fig:3}(a), such that the detection of three photons heralds an entangled GHZ state among the other three~\cite{SLOQC:Rudolph08}, see Supplemental Material. The six input photons are labelled $i_1,...i_6$, and are first initialised in horisontal polarisation. When one, and only one, photon propagates towards each of the heralding outputs $o_4, o_5$, and $o_6$, then the signal output photons $o_1, o_2$, and $o_3$ are left in a three-particle entangled state. Note that the successful implementation of this protocol requires that all six photons are highly indistinguishable from each other. To confirm that this is the case, we measure all 15 cases of pair-wise indistinguishabilities among the six input photons, and find an average indistinguishability of {$0.923{\pm}0.009$} across all combinations, see Fig.~\ref{fig:3}(b).

The heralded generation of the entangled quantum state requires that no more than one photon is detected at each heralding output. For example, without number-resolution, a pattern with four photons among the three heralding spatial trajectories can not be discerned from another pattern with an exact number of three photons. In such cases, the state produced at signal outputs is not solely the target GHZ state, but it also contains other components with a different number of photons. Therefore, only non-heralded (post-selected) states are generated in the absence of number-resolving detection. Our implementation makes use of pseudo photon-number resolution (PPNR) at {every heralding output, $o_{4,h}, o_{4,v}, o_{5,h}, o_{5,v}, o_{6,h}, o_{6,v}$}, by further splitting each of them into two new outputs and SNSPDs; where $h$ and $v$ denote horisontal and vertical polarisation, respectively. {Therefore, we use 18 SNSPDs in total: six detectors to cover both polarisations of the three signal outputs, and twelve detectors for implementing the polarisation and pseudo number-resolved measurement of the three heralding outputs.} At each PPNR stage (six in total), we condition a measurement such that one of the detectors clicks and the other one does not, which performs the pseudo number resolution of one, and no more, photon. The heralded generation of the target state occurs then by imposing that each of the heralding stages measures at most one photon. 

	\begin{figure}[htp!]
		\centering
		\includegraphics[width=.4\textwidth]{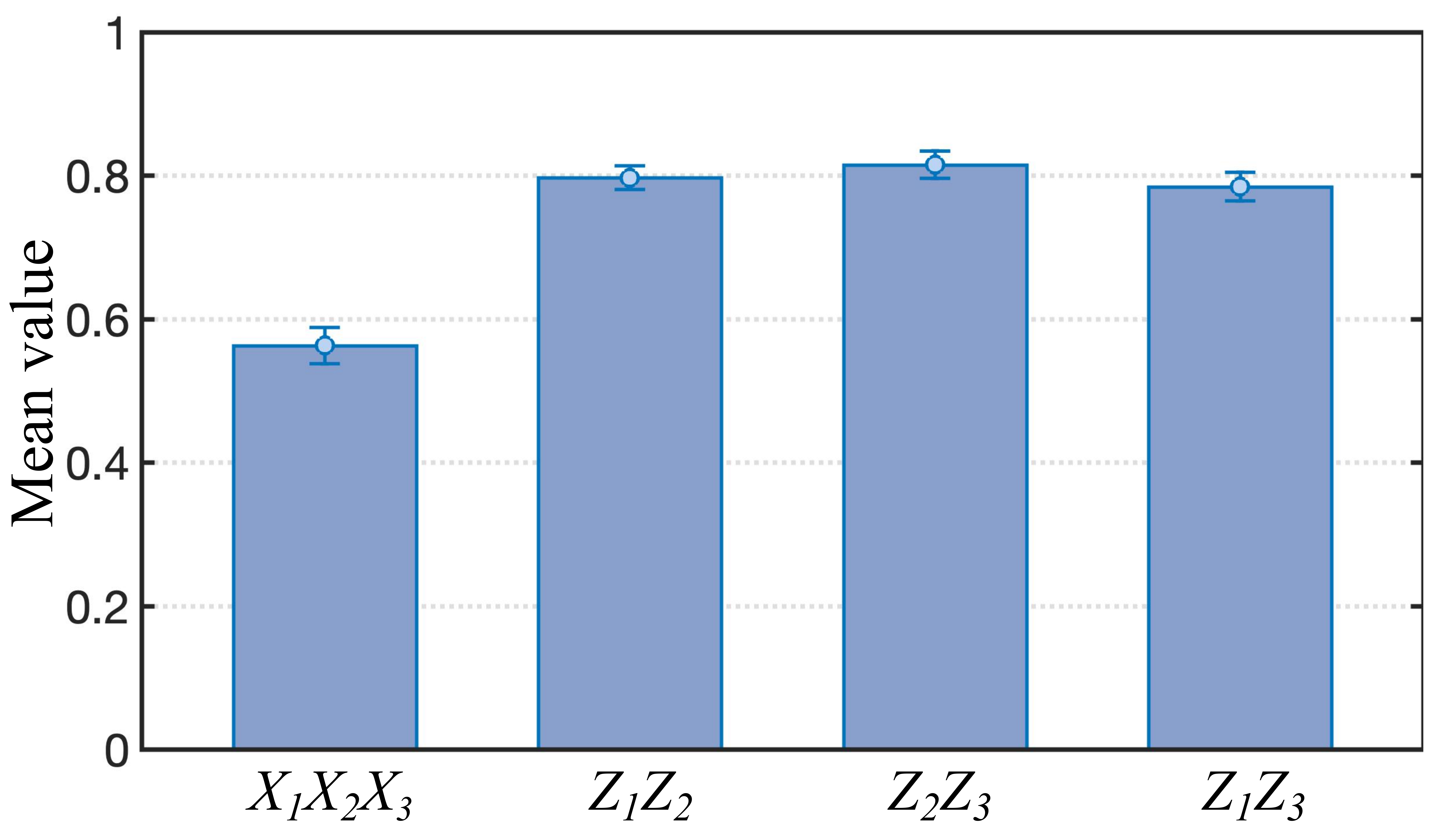}\vspace{0mm}
		\caption{\textbf{Entanglement witness.} Measured mean values of the observables forming $\mathcal{W_\text{GHZ}}$. We obtain $\langle X_1 X_2 X_3\rangle{=}0.5629{\pm}0.0252$, $\langle Z_1 Z_2\rangle{=}0.7971{\pm}0.0166$, $\langle Z_2 Z_3\rangle{=}0.8151{\pm}0.0191$, and $\langle Z_1 Z_3\rangle{=}0.7846{\pm}0.0199$.}
	\label{fig:4}
	\end{figure}
	
	\begin{figure*}[htp!]
		\centering
		\includegraphics[width=.98\textwidth]{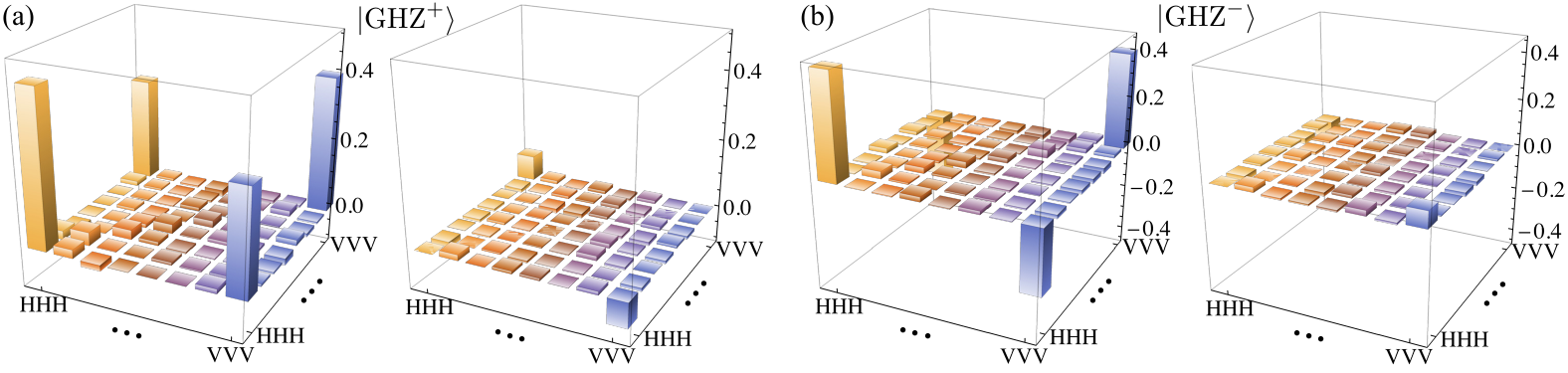}\vspace{0mm}
		\caption{\textbf{Quantum state tomography of the heralded GHZ states.} (a), (b)  Real (left) and imaginary part (right) of the reconstructed density matrices at the signal outputs $o_1, o_2$, and $o_3$, see Fig.~\ref{fig:3}(a), when the heralding patterns produce either the $|\text{GHZ}^+\rangle$ or $|\text{GHZ}^-\rangle$ state. We collect 27 observables for an overcomplete quantum state tomography. To obtain the mean value of a particular observable, we measure simultaneously $2^3{=}8$ polarisation projections among the three signal outputs, times four heralding polarisation combinations that produce the same target state, times $2^3{=}8$ PPNR configurations for number-resolving the three heralding outputs. That is, a total of $2^3{\times}4{\times}2^3{=}256$ cases of six-photon coincidence patterns collected simultaneously among 18 output SNSPDs reconstruct a single three-qubit observable of one heralded state. Accordingly, 512 six-photon coincidence patterns are measured simultaneously to obtain the same observable for both heralded states. Each observable is measured for 900 s, which produced density matrices with a total of 11038 six-fold coincidence counts for the $|\text{GHZ}^+\rangle$ state, and 11178 six-folds for $|\text{GHZ}^-\rangle$. Accordingly, each heralded state is measured at a rate of {$0.454{\pm}0.004$~Hz} and {$0.460{\pm}0.004$~Hz}, respectively, resulting in a total measured rate of $0.914{\pm}0.006$~Hz.}
	\label{fig:5}
	\end{figure*}

The specific 3-GHZ state generated depends on the polarisation pattern that clicks at the heralding outputs. {For instance, the patterns $\left\{hhh, hvv, vhv, vvh\right\}$, where $ijk$ denotes the condition $o_{4,i},o_{5,j},o_{6,k}$, herald the signal state $|\text{GHZ}^{+}\rangle{=}\left(|000\rangle{+}|111\rangle\right){/}\sqrt{2}$ with an accumulated success probability of $1{/}64$, and $|0\rangle$ ($|1\rangle$) denoting horisontal (vertical) polarisation. Conversely, the complementary patterns $\left\{hhv, hvh, vhh, vvv\right\}$ herald the state $|\text{GHZ}^{-}\rangle{=}\left(|000\rangle{-}|111\rangle\right){/}\sqrt{2}$} with an equal success probability. Accordingly, the total success probability of generating a three-GHZ state with this protocol is $1{/}32$.

In our experiment, we start by measuring the witness $\mathcal{W_\text{GHZ}}{=}{3}\mathbb{I}{/}2{-}X_{1}X_{2}X_{3}{-}\left(Z_1 Z_2{+}Z_2 Z_3{+}Z_1 Z_3\right){/}2$, whose negative value verifies the presence of genuine three-particle entanglement for GHZ states~\cite{witnessGHZ:05}. Figure~\ref{fig:4} displays the measured mean values of the involved observables, from where we obtain $\langle\mathcal{W}_\text{GHZ}\rangle{=}-0.2613{\pm}0.0335$, confirming three-partite entanglement by more than seven standard deviations.

Moreover, given the high countrates of the available six-photon source, we are also able to perform overcomplete three-qubit quantum state tomography at the signal outputs, with all heralding patterns simultaneously. That is, by using 18 SNSPDs, we reconstruct both $|\text{GHZ}^{+}\rangle$ and $|\text{GHZ}^{-}\rangle$ from the measurement of $3^{3}{=}27$ three-qubit observables that result from all combinations of $Z,X,Y$ Pauli matrices among the 3 signal qubits. Figure~\ref{fig:5} shows the reconstructed density matrices of the heralded entangled states, from where we extract a fidelity of {$\mathcal{F}^{+}{=}0.7189{\pm}0.0109$ to the $|\text{GHZ}^{+}\rangle$ state when using the corresponding four heralding conditions, as well as $\mathcal{F}^-{=}0.6995{\pm}0.0116$ to $|\text{GHZ}^{-}\rangle$ by using the respective other four heralding patterns.} Note that small terms are present in the imaginary part of the density matrices, showing that the prepared states contain a small relative phase between the state basis, which can be compensated for via local unitaries. Taking this into account, we obtain fidelities of $\overline{\mathcal{F}}^+{=}0.7278{\pm}0.0106$ and $\overline{\mathcal{F}}^-{=}0.7083{\pm}0.0120$ to the GHZ states $\left(|000\rangle{\pm}e^{i\left(0.04{\times}2\pi\right)}|111\rangle\right){/}{\sqrt 2}$, respectively.

We measure both heralded states at a combined rate of {$0.914{\pm}0.006$~Hz}. This value is expected considering: ${\sim}80\%$ throughput efficiency of the polarisation interferometer (affecting six photons), ${\sim}85\%$ throughput efficiency of the pseudo number-resolving detection setup (three photons), and ${\sim}85\%$ throughput given by multiple fibre mating connections (six photons). Together with a $1{/}32$ success probability of producing both GHZ states, results in an expected rate of {$\left(547{~\text{Hz}}\right){\times}0.8^6{\times}0.85^3{\times}0.85^6{/}32{\sim}1$~Hz}.

\textit{Discussion.---}We have experimentally demonstrated a building block for ballistic and all linear-optical photonic quantum computing: the heralded  three-photon Greenberger-Horne-Zeilinger state. First, we developed a high-rate {($547{\pm}2$~Hz)} source of six photons from a solid-state quantum emitter, with an average pair-wise indistinguishability of {$0.923{\pm}0.009$}. Subsequently, these photons propagated through a polarization-based multimode interferometer, where the pseudo number-resolved detection of three of them heralded the generation of three-GHZ states among the remaining particles. Thanks to the high rate of the generated multiphoton source, we were able to perform three-qubit overcomplete quantum state tomography, reaching fidelities up to {$\mathcal{F}{=}0.7278{\pm}0.0106$}. Moreover, the efficient multiphoton source presented here reached an eight-photon rate of ${15.7\pm0.4}$~Hz, readily enabling the implementation of more complex experiments at scales soon beyond ten photons.

Our results represent a significant advancement in the active field of photonic quantum computing. We anticipate that near-term future improvements will overcome limitations of our present experiment. An important issue will be the realization of photon detectors with true photon-number resolution. Otherwise, the heralding events cannot guarantee the presence of GHZ states with unity probability, even if all other elements are perfect in principle. However, despite these constraints, our efforts to utilize Photon Number Resolving Detectors (PNRDs) remain highly valuable. The utilization of pseudo photon-number resolution enhances the probability of generating heralded GHZ states, conditioned on the heralding event, by excluding undesired states. This represents a significant advancement beyond non-PNRDs, which are theoretically unsuitable for heralded GHZ protocols. Our endeavors undeniably push the boundaries of multiphoton entanglement generation in a heralded manner, setting a benchmark for future experiments in this research direction. We foresee that further experiments employing true PNRDs~\cite{cheng2023100} may open up new horizons in this field.

Photonic cluster states have gained considerable attention owing to their remarkable properties and versatility in quantum science. Since the inception of one-way quantum computing~\cite{walther2005experimental}, progress in cluster states has expanded significantly. These achievements range from basic cluster state preparation to blind delegated computation~\cite{barz2012demonstration}, as well as diverse applications beyond~\cite{li2019experimental,shettell2020graph}. Particularly promising is the use of integrated chips to implement complex interferometers and programmable circuits for cluster state preparation, which has led to more practical demonstrations~\cite{adcock2019programmable,huang2023chip,ciampini2016path,vigliar2021error}. These advancements have primarily been accomplished via non-deterministic, post-selected cluster states.
The heralded generation of multiphoton entangled states not only offers a building block for scalable, ballistic universal quantum computation but also provides new insights into relevant applications and fundamental research. For example, the use of genuine multipartite sources, achieved through the combination of bipartite resources and post-selection, for certifying multipartite nonlocality has been a notorious strategy in these experimental endeavors. Their validity has been questioned due to selection bias introduced by post-selection~\cite{blasiak2021safe}. However, the generation of multiphoton entangled states in an event-ready manner may pave the way to close the post-selection loophole in future experiments~\cite{huang2022experimental,cao2022experimental}. 
While heralded entanglement generation has previously been realized only for bipartite states~\cite{hamel2014direct,barz2010heralded}, our experiment marks a steady progression towards multipartite systems. This opens up a plethora of quantum photonics protocols that were previously experimentally inaccessible and removes the constraints of outcome post-selection.

\vspace{3mm}

\textit{Note added.---}During the writing of this manuscript, we became aware of related works~\cite{h3ghz_Pan_PRL, Maring2024}.

\vspace{3mm}

\textit{Acknowledgments---}This research was funded in whole, or in part, from the European Union’s Horizon 2020 and Horizon Europe research and innovation programme under grant agreement No 899368 (EPIQUS) and No 101135288 (EPIQUE), the Marie Skłodowska-Curie grant agreement No 956071 (AppQInfo), and the QuantERA II Programme under Grant Agreement No 101017733 (PhoMemtor); from the Austrian Science Fund (FWF) through Quantum Science Austria (COE1), BeyondC (Grant-DOI 10.55776/F71) and Research Group 5 (FG5); from the Austrian Federal Ministry for Digital and Economic Affairs, the National Foundation for Research, Technology and Development and the Christian Doppler Research Association.

\bibliography{biblio_h3ghz}
\end{document}